\documentclass[twocolumn,showpacs,preprintnumbers,amsmath,amssymb]{revtex4}
\usepackage{graphicx}

\newcommand{\ie}{{\em i.e.~}}

\begin{document}

\title{Universality of One-Dimensional Heat Conductivity}

\author{Trieu Mai and Onuttom Narayan}
\affiliation{
Department of Physics, University of California, Santa Cruz, CA 95064, USA}

\date{\today}

\begin{abstract}
We show analytically that the heat conductivity of oscillator chains
diverges with system size $N$ as $N^{1/3},$ which is the same as
for one-dimensional fluids. For long cylinders, we use the hydrodynamic
equations for a crystal in one dimension. This is appropriate for
stiff systems such as nanotubes, where the eventual crossover to a
fluid only sets in at unrealistically large $N.$ Despite the extra
equation compared to a fluid, the scaling of the heat conductivity
is unchanged. For strictly one-dimensional chains, we show that the
dynamic equations are those of a fluid at {\it all\/} length scales
even if the static order extends to very large $N.$ The discrepancy
between our results and numerical simulations on Fermi-Pasta-Ulam
chains is discussed.
\end{abstract}

\pacs{}

\maketitle 

\section{Introduction} 

When a small temperature gradient is applied across a system, the
heat current that flows through it is expected to follow Fourier's
law, $j = -\kappa\nabla T$~\cite{foot0}.  However, for one-dimensional
systems there are numerous~\cite{foot1} examples of systems where
numerical
simulations~\cite{LLPreview,hatano,grassberger,casati,cipriano,RC,LLP98,LLP03,prosen,hu,wang}
or exact analytical results~\cite{anomaly,dhar,RG} have shown that
Fourier's law breaks down. The thermal conductivity $\kappa$ is
observed to be anomalous, \ie dependent on the system size and
divergent in the thermodynamic limit. Despite the existence~\cite{RG}
of an analytical renormalization group (RG) prediction for the
behavior of $\kappa,$ there is still considerable controversy about
whether the theory applies to chains of stiff oscillators, where
the particles form a lattice to an excellent approximation.  Since
this is the appropriate description for nanotubes, this class of
systems is of great topical interest. An understanding of $\kappa$
also has broader implications, since other transport coefficients
can also be anomalous in low dimensional systems~\cite{RG}.

Numerical~\cite{hatano,grassberger,casati,cipriano,RC,LLP98,LLP03,prosen,hu,wang}
and analytical~\cite{RG,LLP98,wang,prosen,pereverzev,li} studies
show $\kappa\propto N^\alpha$ for various one-dimensional systems,
where $N$ is the size of the system and $\alpha > 0.$ A renormalization
group (RG) analysis~\cite{RG} has shown that for one-dimensional
nonintegrable systems with momentum conservation, $\alpha$ has the
universal value of $1/3,$ and that without momentum conservation
the conductivity is normal ({\em i.e.\/} $\alpha=0$). This result
was derived using the hydrodynamic equations of a normal fluid.
Simulations on one-dimensional hard particle gases have yielded a
wide range of values for $\alpha$~\cite{dhar,casati,grassberger}.
This is because very large system sizes are needed before the large
$N$ behavior is reached, presumably because of the limited phase
space accessed in one-dimensional elastic collisions. This problem
has been circumvented by using a random collision (RC) model~\cite{RC},
where convergence to the asymptotic regime is seen for small $N.$
Recent simulations measuring energy diffusion in hard particle
systems have confirmed~\cite{cipriano} that $\alpha = 1/3.$

In marked contrast to these results, chains of nonlinear oscillators
such as Fermi-Pasta-Ulam (FPU) chains~\cite{fpu} and their variants
consistently show numerical results that disagree with the RG
prediction~\cite{LLP98,LLP03,prosen,hu}. Varying parameters within
a given model gives different values for $\alpha$~\cite{LLP98}, but
they generally lie between 0.37 and
0.44~\cite{LLP98,LLP03,LLPreview,prosen} with occasional exceptions.
This is contrary to the RG prediction, but is compatible with the
result from a mode-coupling analysis that yields $\alpha =
2/5$~\cite{LLP98,wang}. These numerical results have led to suggestions
that there are (at least) two universality classes for heat conduction
in one-dimensional momentum conserving systems~\cite{LLP03,wang},
and that the``fluid-like" RG analysis is not appropriate for
oscillator chains, which are better described as a crystal with
small fluctuations in particle positions.

Although any one-dimensional system will {\em eventually\/} renormalize
to fluid like behavior at sufficiently long length scales, for stiff
oscillator chains the crystalline order persists to very large $N.$
Since the hydrodynamic equations for a crystal are different from
those for a fluid, the large $N$ behavior for such chains could be
different from the fluid prediction.

In this paper, we show that this is not the case: $\alpha$ should
be 1/3 even for stiff oscillator chains.  In this light, the numerical
results on FPU chains are probably due to the well-known difficulties
with equilibrating them~\cite{fpu,berchialla}, necessitating extremely
large $N$ before the asymptotic behavior is seen.

In Section II, we first argue that the earlier RG analysis~\cite{RG}
applies to chains as well as gases. This result is surprising in
view of the fact that the standard hydrodynamic description of
crystals introduces extra degrees of freedom for broken
symmetries~\cite{forster,martin}, and is peculiar to one-dimensional
chains with single-well interparticle interaction potentials. We
then derive the hydrodynamic equations for chains with multiple-well
interaction potentials and for quasi one-dimensional systems such
as tubes~\cite{foot2}.  There are four equations instead of three
for a fluid, because of the extra degree of freedom corresponding
to the broken symmetry. We show that these modified hydrodynamic
equations also yield $\alpha = 1/3.$ In Section III, we elucidate
the discrepancy between numerical results on FPU chains and the
analytical prediction by performing simulations on FPU chains with
collisions, which interpolate continuously between the RC model and
pure (collisionless) FPU chains. As we tune parameters to move
towards pure FPU chains, a new intermediate length scale regime
emerges and increases in range, while leaving the large $N$ scaling
of $\kappa$ unchanged. This corroborates the analytical prediction
that $\alpha = 1/3$ asymptotically.  but shows that this will be
seen at unusually large $N$ for pure FPU chains.  Finally, in Section
IV, we discuss the applicability of these predictions to physical
systems.

\section{1-d Hydrodynamics}

The RG analysis discussed earlier~\cite{RG} uses the hydrodynamic
equations for a one-dimensional fluid~\cite{forster,chaikin}, 
\begin{eqnarray}
\partial_t \rho + \partial_x g & =  & 0, \nonumber \\
\partial_t g + \partial_x (gv) & = & -\partial_x p +
\eta\partial_x^2v, \nonumber \\
\partial_t \epsilon + \partial_x [(\epsilon+p)v] & = & \kappa 
\partial_x^2 T+\eta [(\partial_x v)^2 
+ v\partial_x^2 v] 
\label{normal}
\end{eqnarray}
where $\rho,$ $g$ and $\epsilon$ are the local mass, momentum and
energy densities respectively.  The thermodynamic fields, temperature,
pressure, and velocity are represented by $T$, $p$, and $v$
respectively. There are two transport coefficients, the viscosity $\eta$
and the thermal conductivity $\kappa$.

The RG analysis adds noise terms to the second and third of
Eqs.(\ref{normal}) (the first is an exact identity and therefore
has no noise term). Using the Green-Kubo relations~\cite{green,kubo},
the critical exponents are derived from the equilibrium fluctuations
of a system at constant temperature. This can be done without
approximations by invoking Galilean invariance and the condition
that equal time correlations must obey equilibrium statistical
mechanics.

For a crystal, broken symmetries introduce additional long-lived
hydrodynamic modes~\cite{martin,forster,chaikin}.  For a $d$
dimensional crystal, the extra $d$ broken symmetry degrees of freedom
are associated with~\cite{martin} the $d-1$ transverse sound modes,
which become propagating instead of diffusing, and an extra vacancy
diffusion mode~\cite{martin}.  This vacancy diffusion mode is the
only one that survives for $d=1.$ The number of hydrodynamic equations
is increased from three to four. We are thus led to enquire whether
the results of Ref.~\cite{RG} are altered for a stiff chain, which
can be viewed as a one-dimensional crystal.

Before doing this, we show that for one-dimensional monoatomic
systems, if the interparticle interaction has only one minimum,
vacancies do not exist. The fluid description of Ref.~\cite{RG} is
therefore valid even if the interactions are sufficiently stiff
that the particles are almost on a perfect lattice. This is true
for FPU chains.

This result can be shown on physical grounds: if we remove a single
particle from an otherwise perfect $d$ dimensional crystal, there
is a slight distortion in its vicinity, but a vacancy remains. This
vacancy only moves when a neighboring particle hops into the empty
site, resulting in very slow diffusive motion of vacancies (and
interstitials) through the system. In contrast, if a particle is
removed from a FPU-like one-dimensional chain, its neighbors move
immediately, on microscopic ---\ie non-hydrodynamic --- time scales
to close the gap. Although the reduction in mass density caused by
removing the particle is not eliminated, but only smeared out, this
is what one would expect, since mass density is conserved in
Eqs.(\ref{normal}).  The extra degree of freedom introduced in the
standard hydrodynamic treatment of
crystals~\cite{martin,forster,chaikin,fleming}, distortions of the
broken symmetry, is thus indistinguishable from density fluctuations,
and the fluid description of Eqs.(\ref{normal}) remains valid.

There are two ways in which the argument of the previous paragraph
can break down. First, if the interparticle potential has multiple
minima, it is possible to create a situation where the separation
between adjacent particles is $x_{i+1} - x_i = a + \delta_{i,n} b$
for some $n,$ and the force on all the particles is zero. Here $a$
is the lattice constant and $b$ is the separation between minima
of the interparticle potential, in the simplest case equal to $a$.
The resulting vacancy at site $n$ will diffuse slowly through the
system and is therefore a hydrodynamic mode, distinct from lattice
distortions and momentum which combine to produce longitudinal
sound. Second, if the system is only quasi one-dimensional, for
example a cylinder, vacancies can be introduced as for $d>1.$
Neither of these exceptions applies to FPU chains.

The second exception is, however, applicable to nanotubes.  In view
of their importance, we derive the hydrodynamic equations for a
one-dimensional crystal, modifying Eqs.(\ref{normal}), and show
that the heat conductivity exponent $\alpha$ is {\it still\/} not
changed.  This is because the assumptions that allowed $\alpha$ to
be calculated exactly for a fluid are still valid for the crystal.
(Nanotubes have additional transverse and torsional degrees of
freedom, but we will argue later that these do not affect $\alpha$
either.)

A hydrodynamic theory is formed from the continuity equation of the
conserved quantities and the time derivative of all the broken
symmetries. In the case of the 1-d crystal, the equations are
\begin{eqnarray} 
\partial_t \rho + \partial_x g = 0, \nonumber \\ 
\partial_t g + \partial_x \pi = 0, \nonumber \\  
\partial_t \epsilon + \partial_xj^{\epsilon} = 0, \nonumber \\ 
\partial_t u = v + j^u,
\end{eqnarray}
where $u$ is the displacment and the currents, $\pi$, $j^{\epsilon}$,
and $j^u$, are still to be determined. Galilean invariance demands
the inclusion of the $v$ in the last equation.

Following the methods from~\cite{chaikin} and~\cite{fleming},
constitutive equations can be derived from entropy arguments.  The
first law of thermodynamics for this sytem can be rewritten in terms
of densities,
\begin{equation}
Tds = d\epsilon-\mu d\rho-vdg-hd(\partial_xu),
\end{equation} 
where $s$ is the entropy density and $\mu$ is the chemical potential
per unit mass. A uniform translation cannot alter the energy,
therefore energy can only depend on gradients of $u$. The conjugate
field to $\partial_xu$ is $h = \partial E/\partial(\partial_xu)$
at constant $s, \rho,$ and $g$. The displacement variable also
changes the pressure to
\begin{equation}
p=-\epsilon+Ts+\mu\rho+vg+h\partial_xu.
\end{equation}
Using these two relations and the continuity equations, we can derive
an entropy ``continuity'' equation with source terms,
\begin{eqnarray}
T[\partial_t s + \partial_x(vs+\frac{Q}{T})] = -\frac{Q}{T}\partial_x T 
-(g-\rho v)\partial_x \mu \nonumber\\
+(j^u+v\partial_x u)\partial_x h-(\pi+h-p-gv)\partial_x v,
\label{entropydensity} \\
Q = j^{\epsilon}-\mu(g-\rho v)-\pi v + hj^u - \epsilon v + 
gv^2+hv\partial_x u.
\end{eqnarray}
By integrating both sides of Eq.(\ref{entropydensity}) over a large
volume where the entropy ``current'' $vs+Q/T$ is negligible at
the surface, we have an expression for the time derivative of the
total entropy,
\begin{eqnarray}
T\frac{dS}{dt} = \int dx [-\frac{Q}{T}\partial_x T +
(\rho v-g)\partial_x \mu\nonumber\\
+(j^u+v\partial_x u)\partial_x h 
-(\pi+h-p-gv)\partial_x v].
\label{entropy}
\end{eqnarray}

The constitutive equations for the currents can now be derived by
using the condition of entropy creation, $dS/dt\geq0$~\cite{chaikin}.
The equality holds only when there is no dissipation. Without
dissipation, each term of Eq.(\ref{entropy}) must independently be
zero, giving the reactive terms of the currents:
\begin{eqnarray}
g & = & \rho v, \nonumber \\
j^u_R & = & -v \partial_x u, \nonumber \\
\pi_R & = & p - h + gv, \nonumber \\
Q_R & = & 0, \nonumber \\
j_R^\epsilon & = & (\epsilon+p-h)v.
\end{eqnarray}
The $-v\partial_x u$ term in the displacement current does not
appear in the derivation of crystal hydrodynamics from~\cite{chaikin}
but appears to us to be correct.  With dissipation, entropy is
produced and the currents have dissipative components:
\begin{eqnarray}
j^u_D & = & \Gamma \partial_x h - \lambda \partial_x T, \nonumber \\
\pi_D & = & -\eta \partial_x v, \nonumber \\
Q_D & = & -\kappa \partial_x T - \lambda T \partial_x h, \nonumber \\
j^\epsilon_D & = & -(\kappa-\lambda h)\partial_x T - (\lambda T+\Gamma h)
\partial_x h - \eta v\partial_x v.
\end{eqnarray}
The transport coefficients $\kappa, \eta, \lambda,$ and $\Gamma$
are all positive. In general, there are higher order terms in the
gradient expansion, but we will only retain these first order terms
(which is valid for small gradients). The mass current $g$, does
not have a dissipative term because it is a conserved density itself.

The displacement field introduces two new transport coefficents:
one associated to the relaxation of its conjugate field $h$ and
the second due to the possibility of cross coupling between the
heat current and the displacement field (and the displacement current
and temperature field). This cross coupling is possible because the
field and the current have opposite signs under time-reversal, a
necessary condition for dissipative currents~\cite{forster}. For
this reason, there are no cross couplings with $\pi$. Using these
expressions for the currents in the continuity equations closes our
set of equations,
\begin{eqnarray}
\partial_t \rho + \partial_x g = 0, \nonumber \\
\partial_t g + \partial_x (gv) = -\partial_xp +\partial_xh+
\eta\partial_x^2v + \zeta_g, \nonumber \\
\partial_t \epsilon + \partial_x [(\epsilon+p-h)v] = \kappa 
\partial_x^2 T+\eta [(\partial_x v)^2 
+ v\partial_x^2 v] \nonumber\\
+\Gamma [(\partial_x h)^2+h\partial_x^2 h] 
+\lambda[T\partial_x^2h-h\partial_x^2T]+\zeta_{\epsilon}, \nonumber \\
\partial_t u + (\partial_xu-1)v = \Gamma\partial_x h 
- \lambda\partial_x T+\zeta_u.
\label{dyneq}
\end{eqnarray}
Eqs.(\ref{dyneq}) are the full nonlinear hydrodynamic equations of
the 1-d crystal. Noise, represented by the $\zeta$'s, is introduced
to the last three equations for the RG analysis.

The hydrodynamic modes and their dispersion relations can be solved
for the linearized theory. The three variable normal fluid has three
modes, a heat diffusion mode and two propagating sound
modes~\cite{forster}. The additional displacement variable now
introduces another diffusive mode representing vacancy diffusion.
A vacancy diffusion mode is seen in the longitudinal hydrodynamics
of the three dimensional crystal as well~\cite{chaikin,martin,fleming}.
In these linearized theories, quantities such as the susceptibilities
and correlations can be calculated and Green-Kubo relations can be
derived~\cite{forster}. Although the details of such calculations
are different with four hydrodynamic modes instead of three,
Eqs.(\ref{dyneq}) are clearly still Galilean invariant, and ---
when no temperature gradient is imposed externally --- have equal
time fluctuations that are drawn from the canonical ensemble. As
in Ref.~\cite{RG}, the various critical exponents are therefore
determined and, in particular, $\alpha = 1/3.$

In summary, for one-dimensional momentum-conserving crystals (that
reach local thermal equilibrium), Eqs.(\ref{normal}) describe the
hydrodynamics when vacancies are not possible, as is the case for
FPU chains.  When vacancies are possible, Eqs.(\ref{dyneq}) apply,
but the symmetry arguments that determine $\alpha = 1/3$ for
Eqs.(\ref{normal}) are still valid.  In the next section, we
numerically show the possible source of the discrepancy between
prior numerical results on FPU-like
lattices~\cite{LLPreview,LLP98,LLP03,prosen} and this exact result.

\section{Non-equilibrium Simulations}

Results of past numerical
simulations~\cite{LLPreview,LLP98,LLP03,prosen,hu} consistently
measure an exponent larger than $1/3$ for FPU-like chains. Due to
the exact derivation of the analytical prediction, the good agreement
between this prediction and gas
simulations~\cite{grassberger,casati,cipriano,RC}, and the well-known
problems of convergence in one-dimensional systems~\cite{kpz}, we
conjecture that the FPU simulations have not reached the asymptotic
large $N$ regime. Going to larger and larger system sizes to confirm
this would be prohibitive. Instead, we use a {\it tunable\/} model
which smoothly interpolates between the RC gas and the FPU chain,
where this is easier to see.

We introduce FPU-like springs to the random collision model. As the
springs become stiffer, the system continuously evolves from the
pure RC gas with $\alpha = 1/3$ to the collisionless FPU chain:
from a fluid to a ``crystal''.  As the spring constants are increased,
we observe that an intermediate length scale regime emerges, whereas
the large $N$ scaling of $\kappa$ remains $\sim N^{1/3}$.

For the simulations, we numerically integrate a system of $N$
particles with nearest neighbor interactions. The interaction
potential between every pair of particles is of the generalized FPU
form
\begin{equation}
V(z) = \frac{k_2}{2}z^2+\frac{k_3}{3}z^3+\frac{k_4}{4}z^4+\ldots,
\label{potential}
\end{equation}
where $z = x_{i+1}-x_i$ is the compression (or elongation) of the
spring and $x_i$ is the deviation from the equilibrium position of
the $i^{th}$ particle. Collisions can be neglected in the limit 
when the lattice constant $a$ is large, but for finite $a$ and finite
temperature, the probability of two particles colliding is nonzero.
When two particles come into contact, $z=-a$, they undergo a random
collision as in Ref.~\cite{RC}.

For the simulations, we only use terms up to fourth order in
Eq.(\ref{potential}). In their original work, Fermi, Pasta, and
Ulam~\cite{fpu} noticed that the odd frequency normal modes do not
mix with the even frequency normal modes when the interparticle
potential is even. To avoid this problem and any others that may
arise from this accidental symmetry of an FPU-$\beta$ chain, we use
a finite $k_3$ in all of our chain simulations.

Heat baths are connected to the end particles to maintain a temperature
gradient across the system. The scaling of the conductivity is
determined by measuring the size dependence of the current maintaining
a (small) temperature difference between the baths.  For our numerics,
the leftmost ($i=1$) and rightmost ($i=N$) particles are connected
to Nose-Hoover heat baths~\cite{nose,hoover}. The auxiliary degree
of freedom introduced in the RC model, the transverse momentum,
does not couple to the baths or enter the dynamics except in
collisions, and so is not seen in the pure FPU (collisionless)
limit.

We adopt a convention in which the energy contained in each spring
is symmetrically shared by the two particles attached to it. The
first and last springs are only attached to one particle and a wall,
therefore their entire energy is attributed to the attached particle.
Using this convention for the energy density, the energy current
consists of two parts. The first is an advective part, $j^{adv}(x,t)=\sum_i
e_iv_i\delta(x-x_i(t))$.  The second is piecewise constant and jumps
at each particle with $j_{i+1,i}-j_{i,i-1} = -\dot{e}_i$.  From
this, $j_{i+1,i}=-(1/2)(v_{i+1}+v_i)V^\prime(x_{i+1}-x_i)$, with
$j_{0,1}$ and $j_{N,N+1}$ equal to the energy currents from the
reservoirs to the end particles.  Thus the total current flowing
through the system is
\begin{equation}
jN = \sum_{i=1}^N e_iv_i - \frac{1}{2}\sum_{i=1}^N(x_{i+1}-x_i+a)
(v_{i+1}+v_i)V^\prime(x_{i+1}-x_i).
\label{curr}
\end{equation}
It is possible to show that the time average $<jN>$ is equal to
$-(a/2)\sum_i (v_{i+1}+v_i)V^\prime(x_{i+1}-x_i)$. However, with
collisions it is more useful to keep Eq.(\ref{curr}) in its entirety,
since $x_{i+1}-x_i+a=0$ at collisions, where $V^\prime$ is singular.
In fact, for hard sphere particles, $jN=\sum_{i=1}^Ne_iv_i$.

Numerical integration of this system is complicated by the combination
of collisions and springs. Without springs, the particles travel
freely and a fast event driven simulation can be
implemented~\cite{grassberger}. For a collisionless system, the
equations of motion can simply be integrated numerically by standard
algorithms. For our simulations, the fourth order Runge-Kutta
algorithm is used. Because of the occurrence of collisions, our
algorithm checks for collisions after every trial timestep. If a
collision is seen to have occurred, the trial step is discarded,
and the integrator evolves the system by a smaller stepsize to the
(extrapolated) point at which the two particles collide. Because
of the need for trial steps, integrating systems with collisions
and springs requires substantial amounts of computational time.

\begin{figure}
\begin{center}
\includegraphics[width=3.25in]{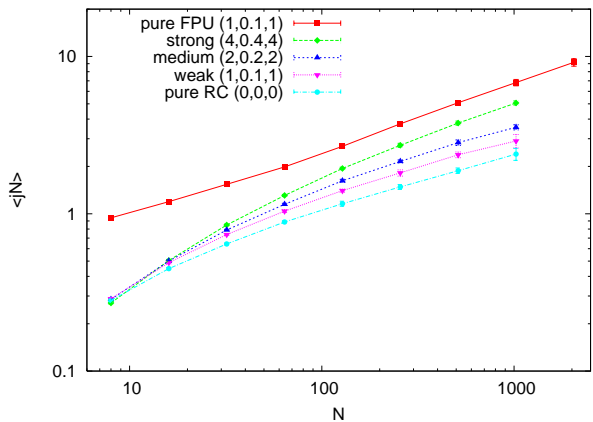}
\caption{Log-log plot of $\langle j N\rangle$ versus $N,$ where $j$
is the time-averaged heat current and $N$ is the number of particles,
for systems with different spring strengths. The spring constants
$(k_2, k_3, k_4)$ of Eq.(\ref{potential}) are specified in the key.
The figure also shows a collisionless FPU chain at the top for
comparison.  The systems without springs or weak springs have slopes
of $\sim 1/3$ at large $N$. The slope increases with spring strength
and is $>0.4$ for the pure FPU system at the largest $N$ shown.
When shown, the error bars are root mean square errors. The plots
are shifted for clarity.}
\label{crossover}
\end{center}
\end{figure}

We use a stepsize of $dt=0.01$ for all simulations and check that
the results do not change for smaller stepsizes. We allow the system
to equilibrate for $\sim 10^8$ steps before measurements are taken.
Each measurement consists of the particle and time averaged current,
where a block of $10^8$ timesteps is used for the time average and
$10^5$ steps separate each block. Each data point in figure~\ref{crossover}
is the mean of many such measurements and the error bars shown are
the root mean square errors of the block measurements.

The parameters in the model are the masses of the particles $m_i$,
the lattice constant $a$, the temperatures of the baths $T_L$ and
$T_R$, and the spring constants $k_n$ in Eq.(\ref{potential}).
Alternating masses are used due to the fast convergence seen in the
pure RC model for such systems~\cite{RC}. In particular, we use a
mass ratio of 2.62 and $m_i=1$ for all odd $i$. The bath temperatures
are 1.2 and 1.0 for the left and right baths, respectively. For
these temperatures, we have checked that the system is in the linear
response regime. With the temperatures of the baths and the masses
of the particles fixed at these values, the collision rate can be
altered by changing the lattice constant and/or spring constants.
We have chosen to fix the lattice constant to $a=1/2$.

\begin{figure}
\begin{center}
\includegraphics[width=3.25in]{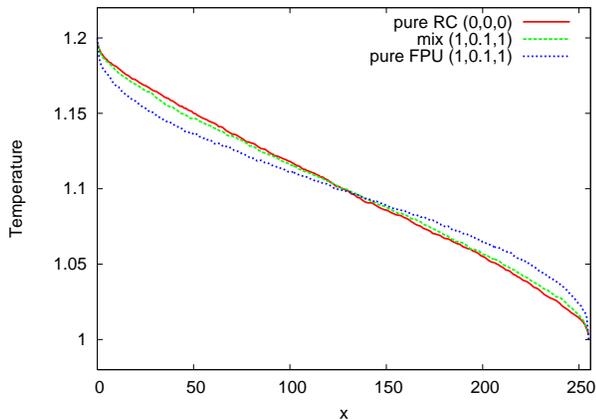}
\caption{The temperature profile for a pure RC gas, a mixed system
with springs and collisions, and a collisionless chain with springs.
The system size $N$ is 256.  The spring constants $(k_2, k_3, k_4)$
of Eq.(\ref{potential}) are $(1, 0.1, 1).$ }
\label{temperature}
\end{center}
\end{figure}

Figure~\ref{temperature} shows the kinetic temperature profile after
steady state is reached for systems of size $N=256,$ showing
increasing curvature as collisions decrease. For the pure FPU system,
the masses of all the particles are chosen to be the same. This is
because with alternating masses, the kinetic temperature for particles
in the pure FPU chain also shows an odd-even oscillation, indicating
that equilibrating the system with Nose-Hoover baths is problematic.
All other systems simulated have alternating masses; as mentioned
in the previous paragraph, this allows the large $N$ limit to be
reached faster.  We verify that the temperature profiles for these
do not oscillate, as shown in Figure~\ref{temperature}.

After the temperature gradient is established, the energy current
is measured. Figure~\ref{crossover} shows a log-log plot of the
current times the number of particles $\langle jN\rangle$ versus
$N;$ the slope of the plot determines $\alpha.$

For no springs, the pure RC model has a slope of $\sim 1/3$, similar
to the earlier results in~\cite{RC}.  This asymptotic value is
reached for relatively small systems ($N\geq128$). As the springs
are turned on, an intermediate length scale regime emerges and the
asymptotic large $N$ regime moves out, but the slope in the large
$N$ regime is {\it still} $1/3.$ For the strongest springs shown
in Figure~\ref{crossover}, the asymptotic regime has moved out of
the range simulated. Figure~\ref{crossover} also shows the results
for a pure FPU chain, which agrees with earlier
results~\cite{LLP98,LLP03,prosen,hu}. The slope of $>0.4$ for this
system is comparable to the strong spring system with collisions,
supporting the assertion that this is an intermediate length scale
phenomenon. As discussed in the previous paragraph, the masses of
all the particles in the pure FPU chain are chosen to be equal.
This is the cause for the different behavior for very small $N,$
where the slope increases with $N$ instead of decreasing.

\section{Discussion}
In the light of the experience with hard particle gases and the
random collision model, it is natural to expect that introducing
extra degrees of freedom would cause FPU chains to show their large
$N$ behavior more rapidly.  In this context, Wang and Li~\cite{wang}
have shown that when one transverse degree of freedom is added to
the FPU chain, the conductivity exponent is numerically seen to be
1/3. It would be interesting to see whether other extensions of the
FPU model can achieve the same result. More ambitiously, it would
be very useful if a criterion could be obtained that would easily
determine numerically whether a system reaches local thermal
equilibrium, allowing one to decide whether the hydrodynamic theory
is applicable.

The study of low dimensional transport is not merely of theoretical
interest; the vast potential applications of real quasi one-dimensional
systems, namely nanotubes, demand an understanding of their physical
properties.  Specifically, research and production of electronic
devices at the micrometer and nanometer scale have shown that carbon
nanotubes may be used in very efficient cooling systems~\cite{collins}.
Experimentally, the thermal conductivity of nanotubes has been found
to be extremely high~\cite{kim} which makes nanotubes obvious
candidates for cooling devices in microelectronics.  This potential
for applications motivates a theoretical investigation of the heat
transport properties of low dimensional systems.
Experimental~\cite{kim,hone} and numerical~\cite{grassberger2,berber}
studies of nanotubes have shown that the phonon contribution to
heat transport is much larger than the electronic contribution, so
that the fact that we have only considered lattice motion in this
paper is not a problem.

The two dimensional nature of the tube allows for the existence of
vacancies.  The tubular shape also introduces other transverse
hydrodynamic modes (a broken symmetry and conserved momentum for
torsional and the two transverse motions), but as mentioned earlier,
these modes do not affect the symmetry arguments, and the heat
conductivity exponent is expected to still be 1/3. The longitudinal
hydrodynamics for this quasi one-dimensional system is the four
component theory, Eqs.(\ref{dyneq}).

There are a few concerns in applying the prediction of $\alpha=1/3.$
First, on short length scales, the phonon motion is ballistic.  The
phonon mean free path in carbon nanotubes is $\sim 1\mu m$~\cite{hone},
and our discussion in this paper would apply to systems that are
larger than this scale. Recent thermal conductivity measurements
have only been made on systems with lengths in this order of
magnitude~\cite{hone}. The unusually high conductivity in such
systems is correctly attributed to the ballistic transport nature
at these scales~\cite{hone}. There have yet to be extensive experiments
on the length dependence of $\kappa$ to our knowledge.  To test our
prediction and measure the actual heat conductivity exponent, length
dependent experiments of long ($\gg 1\mu m$) nanotubes must be done.
Second, even beyond this length scale it is possible (based on what
we have seen for FPU chains) that equilibration is imperfect.
However, this is less likely to be a problem with the extra transverse
modes~\cite{wang} which are present in a nanotube, and in any case,
would only change $\alpha$ slightly. Third, under renormalization,
the system flows to its fixed point only for large $N.$ It is not
clear what limitation this imposes, but the nonlinearity in the
hydrodynamic equations that causes anomalous scaling behavior is
the advective term, whose strength is unity and whose effect should
therefore be seen even for small system sizes~\cite{thank,adv}.

In this paper, we have shown analytically that one-dimensional
chains of particles connected with nonlinear springs have a heat
conductivity that diverges as a function of the chain length $N$
as $\sim N^{1/3},$ which is the same as the earlier result for hard
sphere particles and other fluid systems. For quasi one-dimensional
systems, this result is obtained from the crystalline hydrodynamic
equations. It does {\em not\/} rely on the eventual crossover from
a crystal to a fluid which must happen for any one-dimensional
system.  For Fermi-Pasta-Ulam chains and other examples where the
interparticle potential has a single minimum, we have obtained the
stronger result that the fluid hydrodynamic theory~\cite{RG} is
applicable on all length scales for the dynamics, even when there
is excellent static ordering.  In the light of these analytical
results, the numerical results on FPU chains that show a heat
conductivity exponent of $\sim 0.4$ is probably due to imperfect
equilibration.

\section{Acknowledgments}
We would like to thank Josh Deutsch, Abhishek Dhar, Ram Ramaswamy,
and Sriram Ramaswamy for very useful discussions.


\begin{thebibliography}{}
\bibitem{foot2}
Tubes would introduce many more hydrodynamic modes due to the
transverse degrees of freedom including transverse momentum and
torsional modes. These transverse degrees of freedom need to be
suppressed for our broken symmetry 1-d hydrodynamics to be strictly
applicable, though they should not alter the conductivity exponent.
\bibitem{adv}
Advection takes place in a stiff system even 
though the particles do not move far from their lattice positions.
The collective motion of the particles in a large region can transport 
energy forward, transmitting the energy to neighboring regions 
through interactions, just as in a fluid.
\end{thebibliography}
\end{document}